\documentclass[twocolumn,letterpaper,aps,prc,superscriptaddress,nofootinbib,showpacs,floatfix]{revtex4-1}
\usepackage{graphicx}
\usepackage{comment}
\usepackage{setspace}
\usepackage{amsmath}
\usepackage{amssymb}
\usepackage[textwidth=16cm,textheight=22cm]{geometry}
\usepackage{color}
\usepackage{indentfirst}
\usepackage{xspace}
\usepackage{hyperref}
\usepackage{verbatim}
\usepackage{epstopdf}

\usepackage{lineno}

\begin{document}
\title{Multiplicity dependence of strange and multi-strange hadrons in
p$-$p, p$-$Pb and Pb$-$Pb collisions at LHC energies using Tsallis-Weibull
Formalism}

\author{Pritam~Chakraborty}
\affiliation{Indian Institute of Technology Bombay, Mumbai, 
  India-400076}
  
\author{Tulika~Tripathy}
\affiliation{Indian Institute of Technology Bombay, Mumbai, 
  India-400076}
  
  \author{Subhadip~Pal}
 \affiliation{Indian Institute of Science Education and Research Kolkata , Mohanpur, 
  India-741246}

\author{ Sadhana~Dash }
\email{sadhana@phy.iitb.ac.in}
\affiliation{Indian Institute of Technology Bombay, Mumbai, 
  India-400076}

\begin{abstract}
The transverse momentum ($p_{T}$) distribution of strange hadrons ($K_{S}^{0}$
and $\Lambda$) and multi-strange hadrons( $\Xi$ and $\Omega$) measured in p$-$p, p$-$Pb, and Pb$-$Pb collisions at LHC energies have been studied for different multiplicity classes using Tsallis-Weibull (or q$-$Weibull)
formalism. The distribution describes the measured $p_{T}$  spectra for all multiplicity (or centrality)
classes. The multiplicity dependence of the extracted parameters are studied for the mentioned collisions systems. The $\lambda$ parameter was observed to increase systematically with the collision multiplicity and follows a mass hierarchy for all collision system. This characteristic feature indicates that $\lambda$ can be associated to the strength of collectivity for heavy ion collisions. It can also be related to strength of dynamic effects such as multi-partonic interactions and color reconnections which mimic collectivity in smaller systems. The non-extensive $q$ parameter is found to be greater than one for all the particles suggesting that the strange particles are emitted from a source which is not fully equilibrated.

\end{abstract}

\maketitle

\section{Introduction}
The study of collisions of relativistic heavy ions at RHIC and LHC energies allow us to understand the 
formation of the new phase of matter created at extreme conditions of high temperature and high density. 
The key observables studied in heavy ion collisions to characterize the produced system are often studied
in smaller collision systems such as proton-nucleus (p$-$A) and elementary proton-proton (p$-$p) collisions to disentangle the effects of 
initial and final stages of the collision. The recent  observations in high multiplicity p$-$p and p$-$Pb collisions at LHC energies such as the
enhanced production of strange particles \cite{alicepp, alicenature, ppbcas}, the long-range correlations ("ridge" structure) in the near-side region of the two-particle correlations\cite{cms1,cms2}, hardening of spectra of identified particles which become more pronounced for massive particles\cite{alicepp} and few more are reminiscent of creation of a deconfined medium as produced in A$-$A collisions.

One of the proposed signatures of the formation of this novel state is the enhanced production of strange and multi-strange hadrons\cite{rafelski}. The strange valence quarks are predominantly produced in the parton-parton scattering processes
(like flavor creation and flavor excitation) as they are not present in the initial state of colliding nuclei. The abundance of strange 
partons in the deconfined medium results in the production of more strange particles compared to p$-$p collisions\cite{koch,rafelski}.
The yield of strange particles in central and mid-central collisions has been well described by statistical hadronization models assuming a grand canonical ensemble approach. However, in smaller systems like p$-$p collisions, the relative abundance of 
strange hadrons is reduced compared to A$-$A collisions. This observation was attributed to the effects of canonical suppression which imposes a local conservation of strangeness quantum number in a canonical ensemble formalism while producing strange hadrons\cite{cansup}. However, this mechanism could not explain the strangeness production in peripheral collisions at RHIC where a smaller system is 
expected to be produced\cite{cucu}. The system produced in p$-$A  collisions can be considered to be intermediate  between the 
heavy-ion and the p$-$p collisions in terms of particle multiplicity and volume of the system. Therefore, the study and comparison of particle production in these systems is crucial in order to interpret the effects of canonical formalism and hadronization in general. The recent observation of strangeness enhancement in high multiplicity p$-$Pb and p$-$p collisions indicated towards the progressive 
lifting of canonical suppression with increase in particle multiplicity\cite{ppbcas,alicepp}.
Therefore, it is worth investigating the  transverse momentum ($p_{T}$) distribution of the
produced strange hadrons created in such collisions within different statistical frameworks to 
understand the strange particle production as the integrated effects of dynamics of particle production
from initial stages of collision leaves its footprints in the $p_{T}$ spectra.
Generally, Tsallis statistics is used to describe the $p_{T}$ distribution
in  p$-$p collisions and the flow parameter is introduced in Tsallis
distribution to explain the hadronic spectra in heavy ion collisions to account effects of collectivity\cite{cleyman1,wilk, shukla, pptsallis}. 
Recently it was shown that the Tsallis$-$Weibull (or q$-$Weibull) distribution described the $p_{T}$ spectra in 
p$-$p, p$-$A and A$-$A  collisions for the measured $p_{T}$ range\cite{qweib3}. 
In this work, the Tsallis$-$Weibull (q$-$Weibull) distribution has been used to
describe the multiplicity dependence of strange and multi-strange
hadron production in p$-$p, p$-$Pb and Pb$-$Pb collisions at LHC energies. The parameters which
characterize the distribution are studied as a function of charged
particle density and centrality to understand their evolution with collisions dynamics. This study allows to
extend the q$-$Weibull formalism to identified strange particles in order to gain a clear physical interpretation 
of the associated parameters.
\section{The q-Weibull Distribution}
The natural evolving processes where the dynamic evolution is governed
by fragmentation and sequential branching are described by Weibull
distribution\cite{brown1,brown2}. The production of  hadrons in
non-perturbative domain has an inherent cascade branching
fragmentation and therefore Weibull distribution successfully described
the multiplicity distribution of charged particles in hadron$-$hadron
and leptonic collisions\cite{weib1,weib2}. The Weibull-distribution is incorporated with
Tsallis statistics to obtain q$-$Weibull distribution. 

Mathematically, the q$-$Weibull distribution is obtained by replacing
the exponential factor in Weibull distribution by its equivalent q-exponential\cite{qweib1}.

\begin{equation}
  P_{q}(x; q,\lambda , k ) = \frac{k}{\lambda} \left
    (\frac{x}{\lambda}\right )^{k-1}~ e_{q}^{- (\frac{x}{\lambda})^k}
\end{equation}  

where 
\begin{equation}
  e_{q}^{- (\frac{x}{\lambda})^k}  =  (1 -(1-q) (\frac{x}{\lambda})^k)^{ (\frac{1}{1-q})}
\end{equation}  

As observed in previous analysis of $p_{T}$ spectra of charged particles in
heavy ion collisions\cite{qweib3}, the value of the non-extensive $q$ parameter is related to the deviation of the
system from thermal equilibrium. In general, $q > 1$ value is
attributed to the presence of intrinsic fluctuations in the system
and depends upon the observable being measured. The
$\lambda$ parameter can be associated with the mean $p_{T}$  in hadronic collisions or
the collective expansion velocity of hadrons in heavy ion collisions. The $k$ parameter can be
related to system dynamics associated with the type of collisions.

\section{Analysis and Discussion}

In a previous study \cite{qweib3}, the q-Weibull distribution
was successful in explaining the $p_{T}$  distribution of charged
particles in p$-$p, p$-$A , A$-$A collisions for the measured $p_{T}$ range for most of the beam energies studied.
In this work, the analysis has been extended to identified strange and multi-strange hadrons in these collision systems.
The  $p_{T}$ distribution of strange and multi-strange hadrons in different multiplicity classes were fitted with 
q-Weibull function in p$-$p\cite{alicepp}, p$-$Pb\cite{ppbcas,ppbks},and Pb$-$Pb\cite{kshortprl,alicepbpb} collisions at beam energies of 
7 TeV,  5.02 TeV, and 2.76 TeV, respectively as measured by ALICE experiment.

\begin{figure*}
\begin{center}
\includegraphics[scale=0.8]{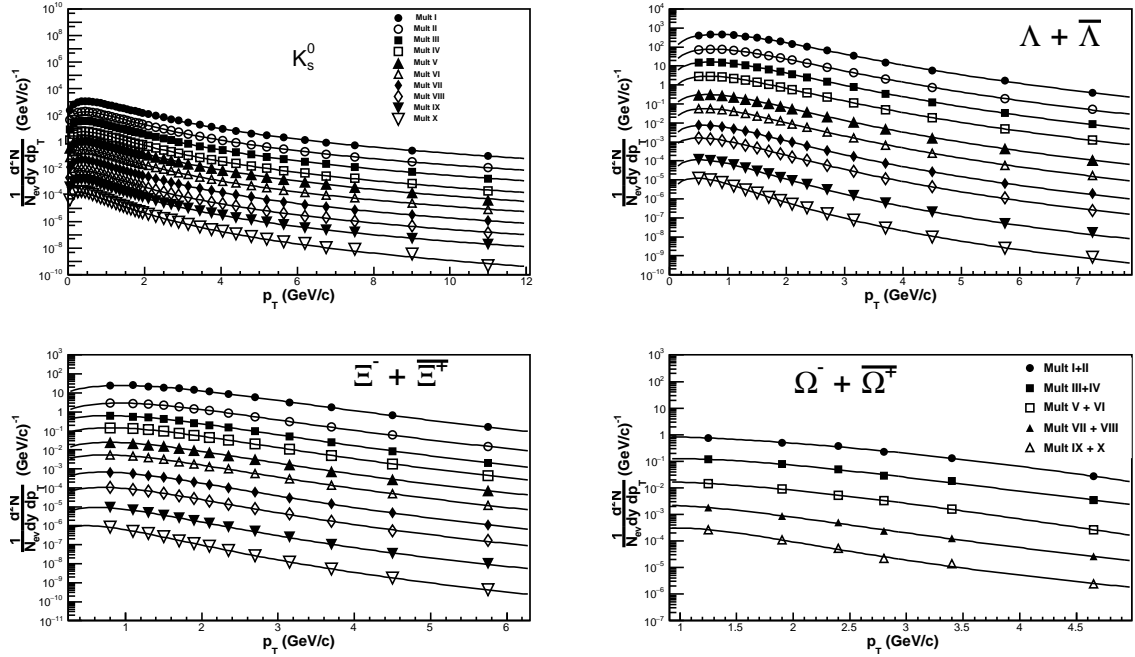}
\caption{(Color online) $p_{T}$ distribution of strange and multi-strange hadrons in p$-$p collisions at $\sqrt{s}$ =  7 TeV for 
$|\eta| < 0.8$ as measured by ALICE  experiment at LHC \cite{alicepp}. The solid lines are the $q$-Weibull fits to the data points. 
The data points are  properly scaled for visibility}
\label{f1}
\end{center}
\end{figure*}

\begin{figure}
\includegraphics[scale=0.4]{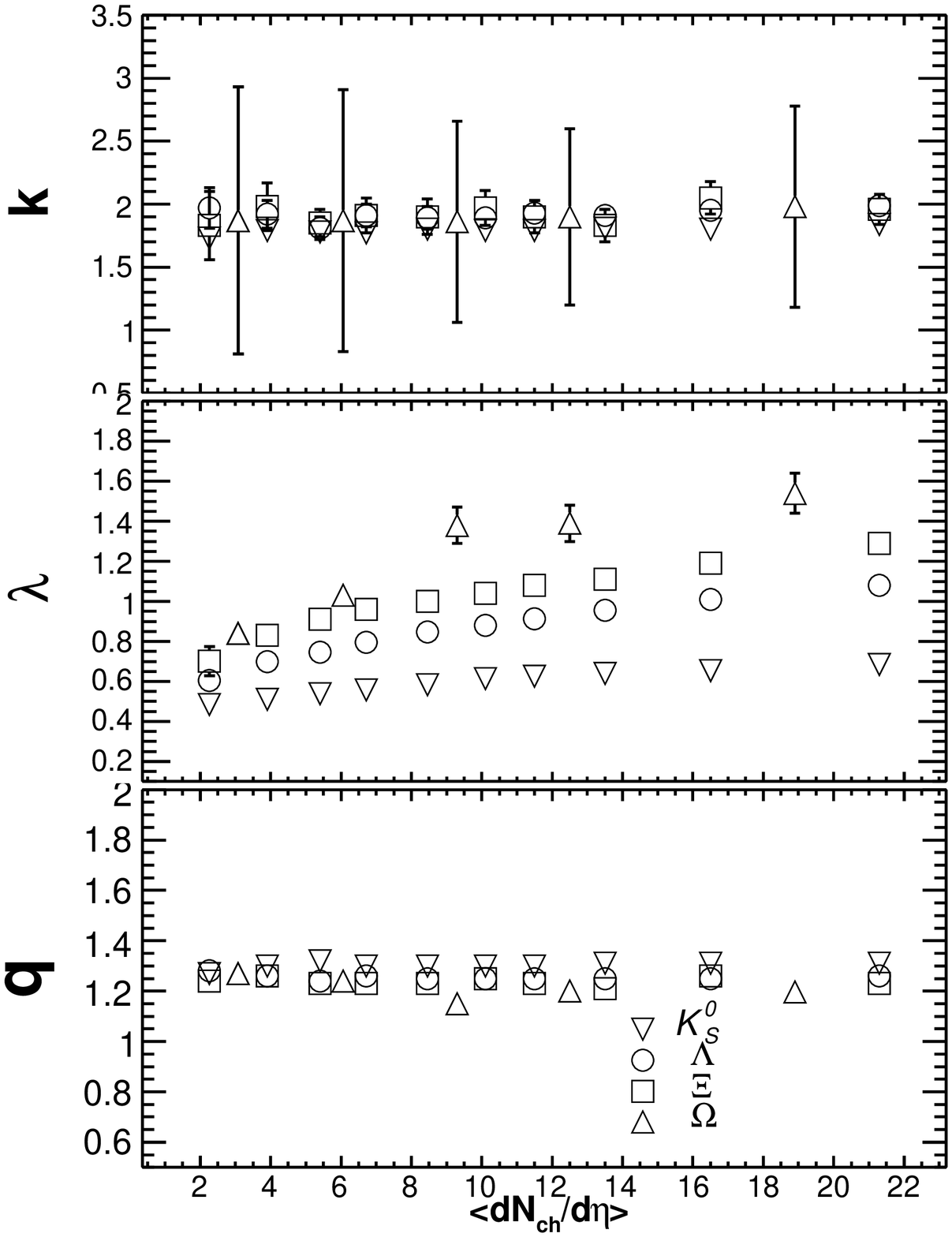}
\caption{(Color online) Variation of $k$ (Upper panel), $\lambda$ (middle panel) and $q$ (Lower panel) as 
a function of $\langle dN_{ch}/d\eta \rangle$ for p$-$p collisions at $\sqrt{s}$ = 7 TeV }
\label{f2}
\end{figure}

Figure~\ref{f1}  shows the invariant yield  of $K^{0}_{S}$, $\Lambda^{0}$, $\Xi^{\pm}$ and $\Omega^{\pm}$ as a function 
of $p_{T}$ for p$-$p collisions at $\sqrt{s}$ = 7 TeV for ten different multiplicity classes. The definition of multiplicity classes and the 
associated  values of ${\langle dN_{ch}/d\eta \rangle }_{\eta < 0.5}$  based on measurement by ALICE experiment\cite{alicenature} is shown in Table \ref{tab1}.
The spectra is  observed to be nicely described by q$-$Weibull function for all multiplicity classes as shown by solid lines in the figure. 
The evolution of values of the extracted  fit parameters with respect to the mean charged particle density ($\langle dN_{ch}/d\eta \rangle$) 
corresponding to different multiplicity classes is shown in Figure~\ref{f2}. The values of the extracted fit parameters together with 
the $\chi^{2}/ndf$ are shown in Table \ref{table2}.  
The $q$ parameter is observed to be  greater than one for the measured strange and multi-strange hadrons for all
multiplicity classes. These values are consistent with the non-equilibrium scenario present in p$-$p collisions.
The $\lambda$ parameter which is related to the mean $p_{T}$  shows a gradual decrease from higher multiplicity classes 
to lower ones. It can also be noted that the values are higher for massive 
hadrons compared to the less massive ones. This  observation seems interesting and is  reminiscent of the onset of collective motion as observed in heavy ion collisions. However, other mechanisms like increase of multi-partonic interactions and color reconnections in final stages as predicted by fragmentation models \cite{dipsy,rope,pythia8} can play an important role in this observation. The $k$ value which was predicted to be related to the dynamical processes contributing to non-equilibrium scenario (dominant in initial stages) is almost constant for the selected hadrons. However, the values are systematically higher for events with higher multiplicity.
\begin{table}
\centering
\begin{tabular}{c c c c }
\hline
Multiplicity  &  $\langle dN_{ch}/d\eta \rangle $ & Multiplicity   & $\langle dN_{ch}/d\eta \rangle$ \\
class & &class&\\
\hline 
\hline
I &  21.5 $\pm$ 0.6   & VI  &  8.45 $\pm$ 0.25 \\
II  &  16.5 $\pm$ 0.5 & VII & 6.72 $\pm$ 0.21\\
III &  13.5 $\pm$ 0.4 & VIII  & 5.40 $\pm$ 0.17\\
 IV  & 11.5 $\pm$ 0.3  & IX & 3.90 $\pm$ 0.14\\
 V   &  10.1 $\pm$ 0.3  & X & 2.26 $\pm$ 0.12\\
\hline 
\end{tabular}
\caption{ \label{tab1} The values of  ${\langle dN_{ch}/d\eta \rangle }_{\eta < 0.5}$  for different multiplicity classes for p$-$p collisions at $\sqrt{s}$ = 7 TeV \cite{alicenature}.}
\end{table}

\begin{figure*}
\centering
\includegraphics[scale=0.8]{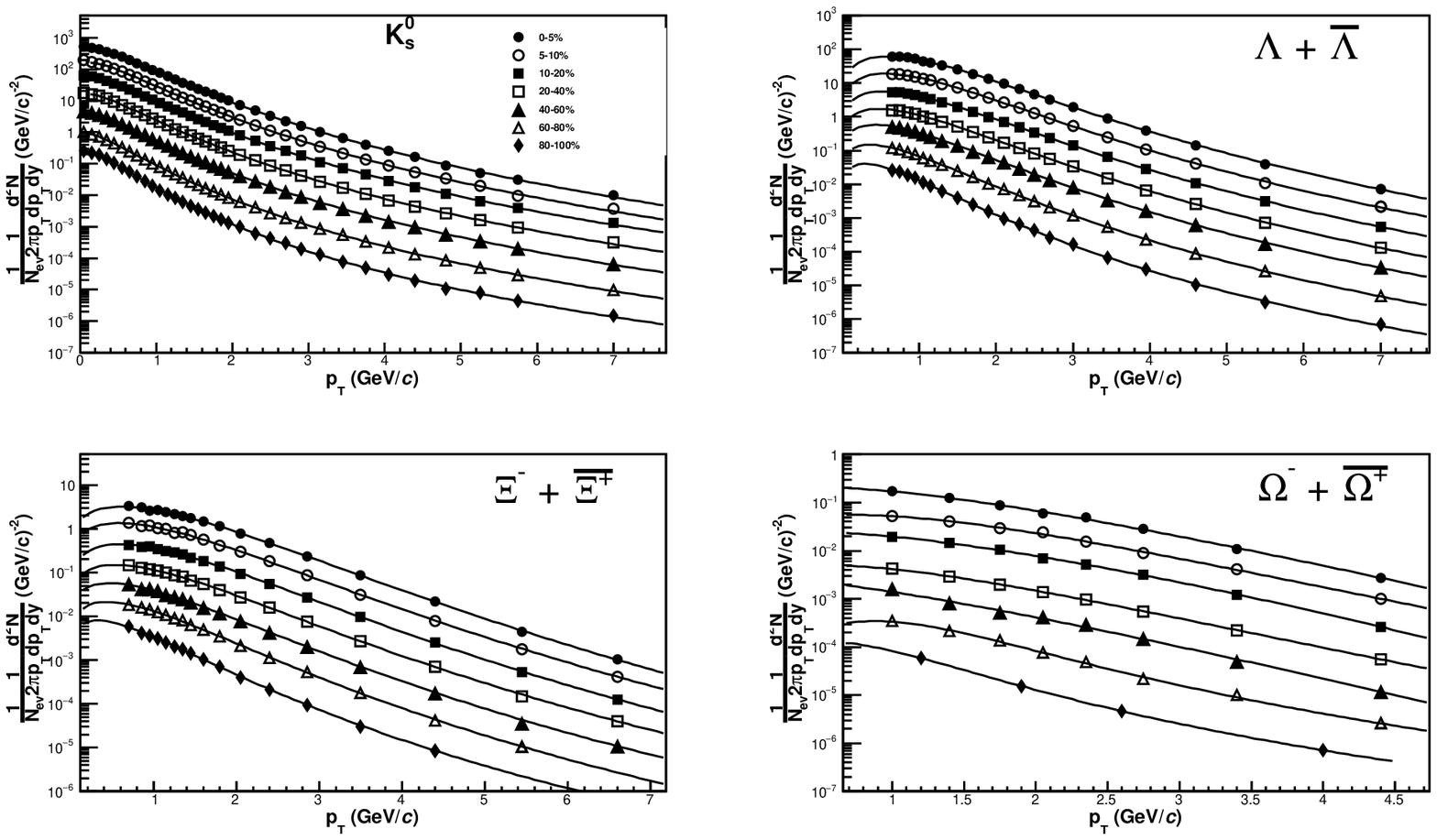}
\caption{(Color online) $p_{T}$ distribution of strange and multi-strange hadrons in different multiplicity classes  
for p$-$Pb  collisions at 5.02 TeV for $|\eta| < 0.8$ as measured by ALICE experiment at LHC \cite{ppbcas,ppbks}. 
The solid lines are the $q$-Weibull fits to the data points. The data points are  properly scaled for visibility. }
\label{f3}
\end{figure*}

\begin{figure}
\includegraphics[scale=0.4]{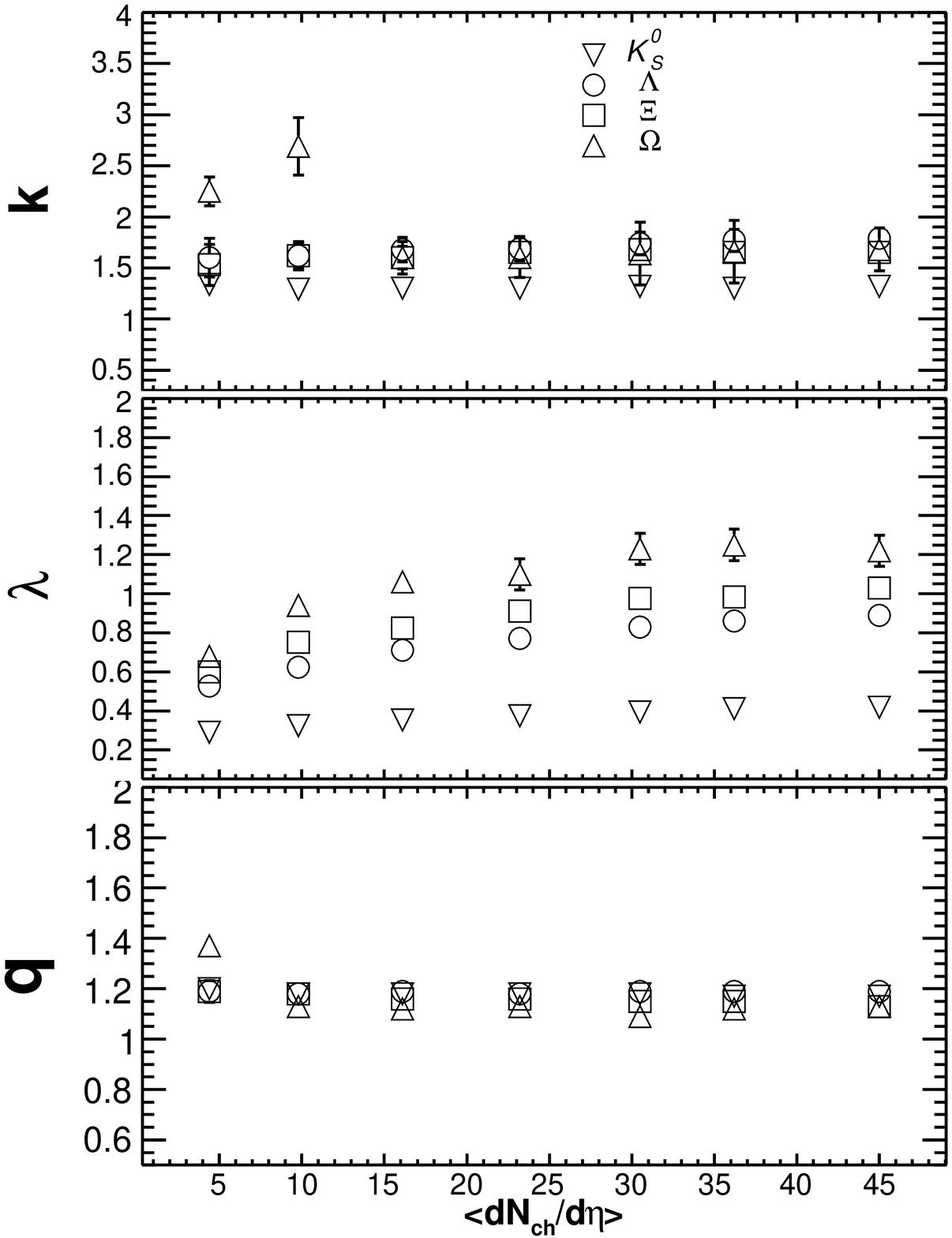}

\caption{(Color online) Variation of $k$ (Upper panel), $\lambda$ (middle panel) and $q$ (Lower panel) as 
a function of $\langle dN_{ch}/d\eta \rangle$ for p$-$Pb collisions at $\sqrt{s}$ = 5.02 TeV }
\label{f4}
\end{figure}

\begin{figure*}
\centering
\includegraphics[scale=0.8]{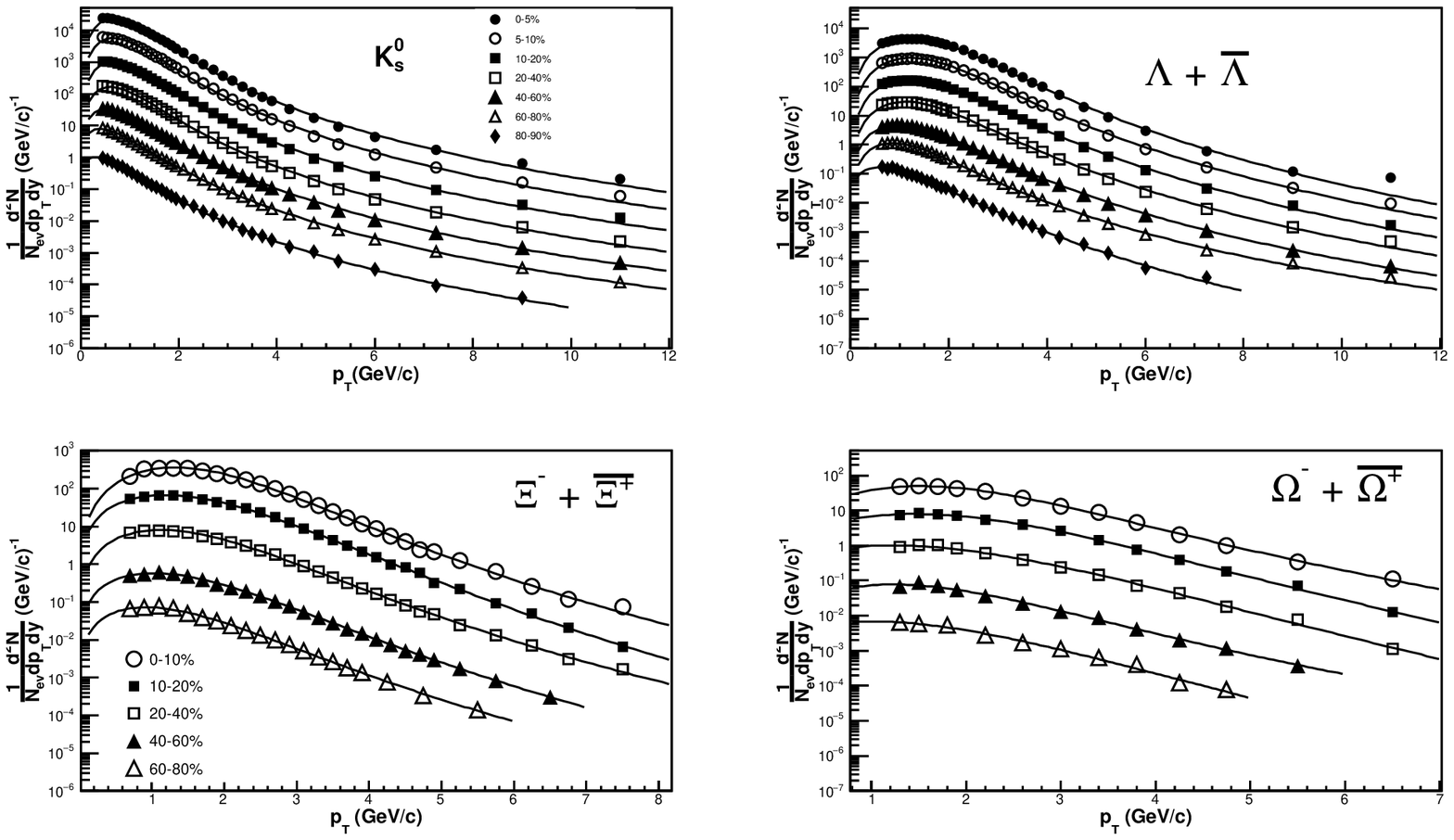}
\caption{(Color online) $p_{T}$ distribution of strange and multi-strange hadrons for various centrality classes  
in Pb$-$Pb  collisions at $\sqrt{s_{NN}}$ = 2.76 TeV for $|\eta| < 0.8$ as measured by ALICE experiment at LHC \cite{alicepbpb,kshortprl}. 
The solid lines are the $q$-Weibull fits to the data points. The data points are  properly scaled for visibility. }
\label{f5}
\end{figure*}

\begin{figure}
\includegraphics[scale=0.4]{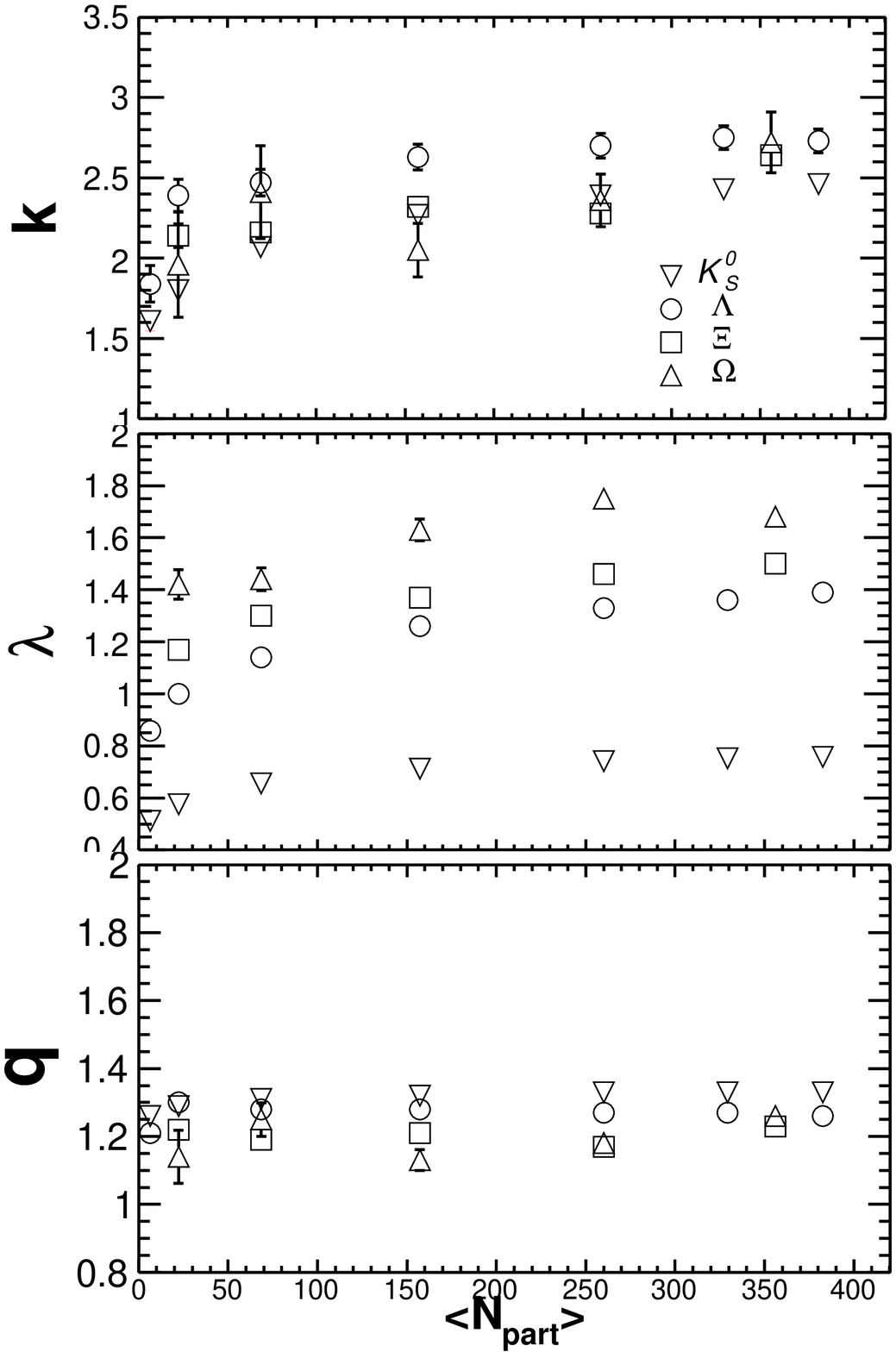}
\caption{(Color online) Variation of $k$ (Upper panel), $\lambda$ (middle panel) and $q$ (Lower panel) as 
a function of $\langle N_{part}$ for Pb$-$Pb collisions at $\sqrt{s_{NN}}$ = 2.76 TeV }
\label{f6}
\end{figure}

Figure~\ref{f3} shows the description of $p_{T}$ spectra for strange
and multi-strange hadrons in p$-$Pb collisions at 5.02 TeV \cite{ppbcas, ppbks} for different
centrality classes. The variation of $q$-Weibull parameters as a
function of $\langle dN_{ch}/d\eta \rangle$ is shown in Fig~\ref{f4}. The $p_{T}$ distribution
of  $K^{0}_{S}$, $\Lambda^{0}$,  $\Xi^{\pm}$ and $\Omega^{\pm}$ in Pb$-$Pb
collisions at 2.76 TeV \cite{alicepbpb} in different centrality
classes is shown in Fig \ref{f5}. Figure \ref{f6}
shows the  centrality ($N_{part}$, number of participating nucleons in a collision)  dependence of parameters for different strange
hadrons.  Table \ref{table3} and Table~\ref{table4} shows the value of
extracted parameters for identified strange particles in p$-$Pb and Pb$-$Pb collisions, 
respectively. As can be observed from the figures, q$-$Weibull fits provide an
excellent description of the data for both the collision systems for a broad
range of centrality classes. The $\lambda$ values show a systematic
increase from peripheral to central collisions for all the studied strange hadrons. The values are also consistently higher for massive hadrons compared to the lesser ones in both the systems. This observation is consistent with the presence of collective motion in heavy ion collisions 
resulting in a common radial flow velocity (as per transverse expansion scenario in hydrodynamic evolution) which increases with centrality
and shows a mass hierarchy. This observation is important and thus $\lambda$ can be related to collective velocity of particles. 
The parameter $k$ shows a slight decrease as we move from central to
peripheral collisions in both Pb$-$Pb and p$-$Pb collisions. This indicates that the value of  $k$ parameter increases with onset of 
processes leading to non-equilibrium conditions.
The deviation from local equilibrium can be quantified by the non-extensive parameter, $q$. 
The $q$ values show a deviation from one which  decreases slightly from central to 
peripheral collisions. A detailed investigation of other identified particles in terms of system 
size and beam energies is required to obtain a clear and consistent interpretation of the parameters.

\begin{table*}
\centering
\begin{tabular}{c c c c c c}
\hline
$ $  &   $\langle dN_{ch}/d\eta \rangle$  & $k$           & $\lambda$   & $q$    &  $\chi^{2}/ndf$ \\
\hline 
\hline
$K^{0}_{S}$   &  21.3  $\pm$  0.6 & 1.84 $\pm$ 0.02 &  0.68 $\pm$ 0.007   & 1.31 $\pm$ 0.005  & 0.8 \\
                      & 16.5   $\pm$  0.5 &  1.80 $\pm$ 0.02 & 0.65 $\pm$  0.007 & 1.31 $\pm$  0.005& 0.54\\
                      & 13.5  $\pm$  0.4 &  1.81  $\pm$ 0.02& 0.64  $\pm$ 0.006& 1.31 $\pm$  0.004& 1.3\\
                  & 11.5  $\pm$ 0.3 &  1.79 $\pm$ 0.02& 0.62 $\pm$  0.006& 1.30 $\pm$ 0.004& 1.65\\
                 & 10.1  $\pm$ 0.3 &  1.79 $\pm$ 0.02&0.61$\pm$   0.006& 1.30$\pm$   0.004& 2.03\\
                 &  8.45  $\pm$ 0.25 & 1.80 $\pm$ 0.02& 0.58 $\pm$  0.005& 1.30 $\pm$  0.004& 1.09\\
           &  6.72  $\pm$ 0.21 & 1.77 $\pm$ 0.02& 0.55 $\pm$  0.005& 1.30 $\pm$  0.004& 0.69\\
       &   5.40  $\pm$ 0.17&  1.78 $\pm$ 0.02& 0.54 $\pm$  0.001& 1.32 $\pm$  0.001& 1.7\\
       &  3.90  $\pm$ 0.14 &  1.79 $\pm$ 0.02 & 0.51 $\pm$  0.005 & 1.30 $\pm$  0.005& 0.9\\
       &  2.26  $\pm$ 0.12 & 1.73 $\pm$  0.03& 0.48$\pm$   0.006& 1.27$\pm$   0.005&2.13\\
\hline 
\hline
$\Lambda + \overline{\Lambda}$  &  21.3  $\pm$ 0.6 & 1.99 $\pm$ 0.08 &1.082 $\pm$ 0.02& 1.26 $\pm$  0.019& 0.18\\
                  & 16.5 $\pm$ 0.5 &  1.95  $\pm$ 0.08& 1.00  $\pm$ 0.018& 1.25  $\pm$  0.02& 0.15\\
                  &  13.5 $\pm$ 0.4& 1.91 $\pm$ 0.08& 0.95 $\pm$  0.017& 1.25 $\pm$  0.02&0.10\\
       & 11.5  $\pm$ 0.3 & 1.93 $\pm$ 0.08& 0.91 $\pm$  0.017& 1.25 $\pm$  0.02 &0.14\\
       & 10.1  $\pm$ 0.3 &1.89  $\pm$ 0.08& 0.88  $\pm$  0.017& 1.25 $\pm$  0.02& 0.16\\
       &  8.45  $\pm$ 0.25 &1.89 $\pm$ 0.09& 0.85  $\pm$  0.016& 1.25  $\pm$  0.02& 0.24\\
       &  6.72  $\pm$ 0.21 & 1.91 $\pm$ 0.09& 0.79 $\pm$ 0.016& 1.26 $\pm$  0.02& 0.17\\
       &   5.40  $\pm$ 0.17& 1.81 $\pm$ 0.09& 0.75 $\pm$  0.003& 1.24 $\pm$  0.02& 1.56\\
       &  3.90  $\pm$ 0.14 &  1.92 $\pm$ 0.11& 0.70 $\pm$  0.02& 1.26 $\pm$  0.02& 0.40\\
       &  2.26  $\pm$ 0.12 & 1.97 $\pm$ 0.16& 0.60$\pm$ 0.03&1.28$\pm$ 0.02& 0.430\\
\hline
\hline
$\Xi^{-} + \overline{\Xi^{+}}$ &  21.3  $\pm$ 0.6 & 1.96$\pm$0.12& 1.29$\pm$ 0.03&1.23$\pm$ 0.03&0.74\\
                  & 16.5 $\pm$ 0.5 &2.05$\pm$ 0.13&1.19 $\pm$ 0.02&1.26 $\pm$0.03& 0.38\\
                  &  13.5 $\pm$ 0.4& 1.83$\pm$0.13&1.11$\pm$0.02&1.21$\pm$0.03&0.53\\
       & 11.5  $\pm$ 0.3 &1.90$\pm$0.13&1.08$\pm$0.025&1.23$\pm$0.03&0.17\\
       & 10.1  $\pm$ 0.3 &1.97$\pm$0.14&1.04$\pm$0.026&1.25$\pm$0.03&0.16\\
       &  8.45  $\pm$ 0.25 &1.89$\pm$ 0.14&1.0$\pm$ 0.02&1.23 $\pm$ 0.03&0.35\\
       &  6.72  $\pm$ 0.21 & 1.91 $\pm$ 0.14&0.96 $\pm$0.03&1.23 $\pm$0.03&0.39\\
       &   5.40  $\pm$ 0.17&1.85 $\pm$0.11&0.91 $\pm$0.02&1.23 $\pm$ 0.027&1.10\\
       &  3.90  $\pm$ 0.14 &1.97 $\pm$0.19&0.83 $\pm$0.04&1.26 $\pm$0.034&0.68\\
       &  2.26  $\pm$ 0.12 &1.83 $\pm$ 0.27&0.70 $\pm$ 0.07&1.24 $\pm$0.04&0.34\\
\hline
\hline
$\Omega^{-} + \overline{\Omega^{+}}$ &  18.9  $\pm$ 0.6 & 1.98$\pm$0.8& 1.54$\pm$ 0.1&1.196$\pm$ 0.03&0.74\\
                  & 12.5 $\pm$ 0.5 &1.90$\pm$ 0.7&1.38 $\pm$ 0.09&1.2 $\pm$0.03& 0.38\\
                  &  9.3 $\pm$ 0.4& 1.86$\pm$0.8&1.37$\pm$0.09&1.15$\pm$0.03&0.53\\
       & 6.06  $\pm$ 0.3 &1.87$\pm$0.13&1.03$\pm$0.01&1.24$\pm$0.03&0.17\\
       & 3.08  $\pm$ 0.3 &1.87$\pm$0.14&0.84$\pm$0.026&1.27$\pm$0.03&0.16\\
\end{tabular}
\caption{ \label{table2} The values of $k$, $\lambda$, $q$ and
  $\chi^{2}/ndf$ obtained from the fits of $p_{T}$ distributions using $q$-Weibull
  function in p$-$p collisions at $\sqrt{s}$ = 7 TeV for different multiplicity classes as 
  measured by ALICE experiment at LHC \cite{alicepp, alicenature}.} 
\end{table*}

\begin{table*}
\centering
\begin{tabular}{c c c c c c c}
\hline
$ $  &   Centrality(\%)  & $\langle dN_{ch} /d\eta \rangle$& $k$           & $\lambda$   & $q$    &  $\chi^{2}/ndf$ \\
\hline 
\hline
$K^{0}_{S}$  &  0-5\% & 45 $\pm$ 1 & 1.32 $\pm$ 0.03 & 0.42 $\pm$ 0.007   & 1.17 $\pm$ 0.008  & 0.4 \\
       & 5-10\%  & 36.2 $\pm$ 0.8 &  1.30 $\pm$ 0.03 & 0.41 $\pm$  0.007 & 1.17 $\pm$  0.008& 0.5\\
       &  10-20\% & 30.5 $\pm$ 0.7 &  1.32  $\pm$ 0.03& 0.39  $\pm$ 0.007& 1.18 $\pm$  0.008& 0.3\\
       & 20-40\% & 23.2 $\pm$ 0.5  & 1.3 $\pm$ 0.03& 0.37 $\pm$  0.007& 1.18 $\pm$ 0.009& 0.2\\
       & 40-60\%  & 16.1 $\pm$ 0.4  &  1.3 $\pm$ 0.03&0.35$\pm$   0.007& 1.18$\pm$   0.009& 0.1\\
       &  60-80\% & 9.8 $\pm$ 0.2  &1.29 $\pm$ 0.03& 0.32 $\pm$  0.007& 1.18 $\pm$  0.009& 0.1\\
       &  80-100\% & 4.4 $\pm$ 0.1  & 1.34$\pm$ 0.04 &0.29 $\pm$  0.007& 1.20 $\pm$  0.009& 0.3\\
\hline 
\hline
$\Lambda + \overline{\Lambda}$  & 45 $\pm$ 1 &  0-5\%& 1.79$\pm$ 0.10 &0.88 $\pm$ 0.03& 1.19 $\pm$  0.02& 0.1\\
       & 5-10\% & 36.2 $\pm$ 0.8 &  1.77  $\pm$ 0.11& 0.86  $\pm$ 0.03& 1.25  $\pm$  0.02& 0.1\\
        &  10-20\%  & 30.5 $\pm$ 0.7 & 1.74 $\pm$ 0.11& 0.83 $\pm$  0.03& 1.25 $\pm$  0.02&0.05\\
       & 20-40\% & 23.2 $\pm$ 0.5 &1.69 $\pm$ 0.12& 0.77 $\pm$  0.03& 1.25 $\pm$  0.023 &0.05\\
       & 40-60\% & 16.1 $\pm$ 0.4 &1.68  $\pm$ 0.12& 0.71  $\pm$  0.03& 1.25 $\pm$  0.02& 0.1\\
       &  60-80\% & 9.8 $\pm$ 0.24 &1.62 $\pm$ 0.14& 0.62  $\pm$  0.04& 1.25  $\pm$  0.03& 0.1\\
       &  80-100\% & 4.4 $\pm$ 0.1 & 1.60 $\pm$ 0.19& 0.52 $\pm$ 0.03& 1.26 $\pm$  0.03& 0.2\\      	
\hline
\hline
$\Xi^{-} + \overline{\Xi^{+}}$ &  0-5\% & 45 $\pm$ 1 & 1.65$\pm$0.09& 1.03$\pm$ 0.03&1.13$\pm$ 0.02&0.3\\
                  & 5-10\%  & 36.2 $\pm$ 0.8 &1.65$\pm$ 0.09&0.98$\pm$ 0.03&1.15$\pm$0.02& 0.5\\
                  &  10-20\% & 30.5 $\pm$ 0.7 & 1.68$\pm$0.09&0.97$\pm$0.02&1.15$\pm$0.02&0.2\\
       & 20-40\%  & 23.2 $\pm$ 0.5  &1.65$\pm$0.09& 0.91$\pm$0.02&1.16$\pm$0.02&0.1\\
       & 40-60\% & 16.1 $\pm$ 0.4 &1.60$\pm$0.11&0.82$\pm$0.03&1.16$\pm$0.02&0.3\\
              & 60-80\% & 9.8 $\pm$ 0.2 &1.62$\pm$0.12&0.75$\pm$0.034&1.18$\pm$0.02&0.1\\
                     & 80-100\% & 4.4 $\pm$ 0.1 &1.53$\pm$0.2&0.59$\pm$0.05&1.19$\pm$0.04&0.2\\

\hline
\hline
$\Omega^{-} + \overline{\Omega^{+}}$ & 0-5\% & 45 $\pm$ 1  & 1.68$\pm$0.2& 1.22$\pm$ 0.08&1.13$\pm$ 0.03&0.6\\
                  & 5-10\% & 36.2 $\pm$ 0.8 &1.66$\pm$ 0.3&1.25 $\pm$ 0.08&1.12 $\pm$0.03& 0.4\\
                  &  10-20\%  & 30.5 $\pm$ 0.7& 1.64$\pm$0.3&1.23$\pm$0.08&1.09$\pm$0.03&0.2\\
       & 20-40\% & 23.2 $\pm$ 0.5 &1.60$\pm$0.1&1.1$\pm$0.08&1.13$\pm$0.03&0.1\\
       & 40-60\% & 16.1 $\pm$ 0.4  &1.60$\pm$0.1&1.06$\pm$0.03&1.12$\pm$0.03&1.6\\
       & 60-80\%  & 9.8 $\pm$ 0.2 &2.69$\pm$0.7&0.94$\pm$0.03&1.13$\pm$0.03&0.5\\
       &  80-100\% & 4.4 $\pm$ 0.1  &2.25$\pm$0.1&0.68$\pm$0.03&1.37$\pm$0.03&0.16\\
\end{tabular}
\caption{ \label{table3} The values of  $k$, $\lambda$, $q$ and
  $\chi^{2}/ndf$ obtained from the fits of $p_{T}$ distributions using $q$-Weibull
  function in p$-$Pb collisions for different centrality classes at
  $\sqrt{s_{NN}}$ = 5.02 TeV as measured by ALICE experiment at LHC \cite{ppbks,ppbcas}.} 
\end{table*}

\begin{table*}
\centering
\begin{tabular}{c c c c c c c}
\hline  
$  $  &   Centrality(\%)  & $\langle N_{part} \rangle$  &$k$           & $\lambda$   & $q$    &  $\chi^{2}/ndf$ \\
\hline 
\hline
$K^{0}_{S}$  &  0-5\% & 382.7  $\pm$ 3.0& 2.46 $\pm$ 0.05 & 0.76 $\pm$ 0.01   & 1.33 $\pm$ 0.01 & 1.8 \\
       & 5-10\% &329.4 $\pm$ 4.3 &  2.43 $\pm$ 0.05 & 0.75 $\pm$  0.01 & 1.32 $\pm$  0.01& 1.5\\
       &  10-20\% &260.1 $\pm$ 3.8 &  2.39  $\pm$ 0.05& 0.74  $\pm$ 0.01& 1.32 $\pm$  0.01& 2.1\\
       & 20-40\% &157.2 $\pm$ 3.1 & 2.27 $\pm$ 0.05& 0.71 $\pm$  0.01& 1.32 $\pm$ 0.01& 1.2\\
       & 40-60\% &68.6 $\pm$ 2.0  &  2.07 $\pm$ 0.05&0.66$\pm$   0.01& 1.30$\pm$   0.01& 1.1\\
       &  60-80\%&22.3 $\pm$ 0.8 &1.80 $\pm$ 0.05& 0.57 $\pm$  0.01& 1.28 $\pm$  0.01& 0.5\\
       &  80-90\% & 6.4 $\pm$ 4.2 & 1.61$\pm$ 0.06& 0.51 $\pm$  0.01& 1.26 $\pm$  0.01& 1.3\\
 \hline 
\hline
$\Lambda + \overline{\Lambda}$  &  0-5\% & 382.7  $\pm$ 3.0 & 2.73 $\pm$ 0.07 &1.39 $\pm$ 0.02& 1.26 $\pm$  0.01& 1.2\\
       & 5-10\% &329.4 $\pm$ 4.3 &  2.75  $\pm$ 0.07& 1.36  $\pm$ 0.02& 1.27  $\pm$  0.02& 1.3\\
        &  10-20\% &260.1 $\pm$ 3.8  & 2.70 $\pm$ 0.08& 1.33 $\pm$  0.02& 1.27 $\pm$  0.02&1.1\\
       & 20-40\% &157.2 $\pm$ 3.1   & 2.63 $\pm$ 0.08& 1.26 $\pm$  0.02& 1.28 $\pm$  0.02 &1.1\\
       & 40-60\%  &68.6 $\pm$ 2.0 &2.47  $\pm$ 0.08& 1.14  $\pm$  0.02& 1.28 $\pm$  0.02& 0.5\\
       &  60-80\% &22.3 $\pm$ 0.8 &2.39 $\pm$ 0.09& 1.00  $\pm$  0.02& 1.30  $\pm$  0.02& 0.9\\
       &  80-90\% & 6.4 $\pm$ 4.2 & 1.84 $\pm$ 0.13& 0.85 $\pm$ 0.03& 1.21 $\pm$  0.02& 0.1\\      	
\hline
\hline
$\Xi^{-} + \overline{\Xi^{-}}$ &  0-10\%  &356.1 $\pm$ 3.6 & 2.64$\pm$0.05& 1.50$\pm$ 0.01&1.22$\pm$ 0.006&1.4\\
           & 10-20\%& 260.1 $\pm$ 3.9  &2.28$\pm$ 0.04 &1.46 $\pm$ 0.01&1.17 $\pm$0.006& 1.8\\
             &  20-40\% & 157.2 $\pm$ 3.1  & 2.31$\pm$0.05  &1.37$\pm$0.02&1.21$\pm$0.007&2.1\\
       & 40-60\% & 68.6 $\pm$ 2.0  &2.16$\pm$0.05&1.30$\pm$0.01&1.19$\pm$0.01&0.7\\
       & 60-80\% & 22.3 $\pm$ 0.8 &2.13$\pm$0.07&1.17$\pm$0.01&1.21$\pm$0.02&1.6\\
\hline
\hline
$\Omega^{-} + \overline{\Omega^{+}}$& 0-10\% & 356.1 $\pm$ 3.6   & 2.71$\pm$0.18& 1.67$\pm$ 0.03&1.25$\pm$ 0.03&1.1\\
                  & 10-20\% & 260.2 $\pm$ 3.9  &2.36$\pm$ 0.16&1.74 $\pm$ 0.03&1.17 $\pm$0.03& 1.2\\
                  &  20-40\% &157.2 $\pm$ 3.1 & 2.04$\pm$0.16&1.62$\pm$0.041&1.12$\pm$0.03&1.3\\
       & 40-60\%  & 68.9 $\pm$ 2.0  &2.41$\pm$0.28 &1.43$\pm$0.043&1.24$\pm$0.04&2.1\\
       & 60-80\% & 22.3 $\pm$ 0.8 &1.95$\pm$0.32&1.41$\pm$0.056&1.13$\pm$0.05&2.6\\
\end{tabular}
\caption{ \label{table4} The values of $k$, $\lambda$, $q$ and
  $\chi^{2}/ndf$ obtained from the fits of $p_{T}$ distributions using $q$-Weibull
  function in Pb$-$Pb collisions for different multiplicity classes at 
 $\sqrt{s_{NN}}$ = 2.76 TeV as measured by ALICE experiment at LHC \cite{kshortprl,alicepbpb}.} 
\end{table*}
\section{Summary}

The study of production of strange and multi-strange particles in different systems 
have become quite relevant with the recent observation of strangeness enhancement in high multiplicity p$-$p collisions at 
LHC energies. The q$-$Weibull statistics has been applied for the first time to comprehend  the $p_{T}$ distribution of strange and
multi-strange hadrons in different colliding systems (p$-$p , p$-$Pb and Pb$-$Pb ) for a broad range
of multiplicity classes. The $q$-Weibull function successfully describes the $p_{T}$ distribution
for all ranges of $p_{T}$ measured. The evolution of the parameters of the distribution extracted from the 
fit were studied  as a function of multiplicity for all the collision systems. 
The $\lambda$ parameter which was conjectured to be associated with the collective expansion velocity was observed to 
systematically increase with the collision centrality in Pb$-$Pb and p$-$Pb collisions. A mass hierarchy was also observed i.e. the particles 
with higher mass had larger values of $\lambda$. These features are consistent with the presence of 
collectivity(radial flow)  in the medium formed and hence one can attribute $\lambda$ parameter to the collective velocity.
A similar behavior for the variation of $\lambda$  with multiplicity was also observed for p$-$p collisions.
For p$-$p collisions, this parameter can be related to strength of other mechanisms like multi-partonic interactions and 
color reconnections which mimic features of collectivity. The $q$ parameter which is related to the  degree of deviation from 
conditions of thermal equilibrium was almost a  constant with respect to centrality for all collision systems studied. The values consistently 
deviated from one for all the collision systems and were also higher for Pb$-$Pb and p$-$Pb collisions indicating that the system  
from which the strange hadrons are emitted is not  fully equilibrated. 
The values of $k$ shows a slight variation with centrality where the values decrease while moving from high multiplicity 
to lower ones for certain particles. It can be related to certain dynamical features of collision which indicates the presence of
processes which are dominant in initial stages of collision. 
\section{Acknowledgement}

The authors would like to thank the Department of Science and Technology
(DST), India for supporting the present work.

\end{document}